\documentclass[submission, Phys]{SciPost}
\usepackage{dcolumn}
\usepackage{bm}
\usepackage{listings}
\usepackage{ulem}
\usepackage{graphicx}
\usepackage{dsfont}  
\newcommand{\mathbb}[1]{\mathds{#1}}

\begin{document}

\vspace{10pt}
\noindent\rule{\textwidth}{1pt}
\tableofcontents\thispagestyle{fancy}
\noindent\rule{\textwidth}{1pt}
\vspace{10pt}

\newcommand{\veca}{{\vec{a} } }
\newcommand{\vecx}{{\vec{x}}}
\newcommand{\vecX}{{\vec{X}}}
\newcommand{\hatS}{{\hat{S}}}
\newcommand{\la}{\langle}
\newcommand{\ra}{\rangle}
\newcommand{\beq}{\begin{equation}}     
\newcommand{\eeq}{\end{equation}}
\newcommand{\beqa}{\begin{eqnarray}}    
\newcommand{\eeqa}{\end{eqnarray}}
\newcommand{\ben}{\begin{enumerate}}   
\newcommand{\een}{\end{enumerate}}
\newcommand{\bde}{\begin{description}}   
\newcommand{\ede}{\end{description}}
\newcommand{\inv}[1]{{\frac{1}{#1}}}
\newcommand{\inAverage}[1]{{\left\la{#1}\right\ra}}

\begin{center}{\Large \textbf{
Autonomous self-harmonic drift in Langevin dynamics causes a compaction\\ of underlying domain for the density of Stochastic Localization}}
\end{center}

\begin{center}
Ken Sekimoto\textsuperscript{1$\star$} 
\end{center}

\begin{center}
{\bf 1} Laboratoire Mati\`ere et Syst\`emes Complexes, UMR CNRS 7057, Universit\'e Paris Cit\'e,		10 Rue Alice Domon et L\'eonie Duquet, 75013, Paris, France,\\ and Laboratoire Gulliver, UMR CNRS 7083, ESPCI Paris, Universit\'e PSL,\\		10 rue Vauquelin, 75005, Paris, France
\\
${}^\star$ {\small \sf ken.sekimoto@espci.fr}
\end{center}

\begin{center}
\today
\end{center}

\section*{Abstract}
{\bf
Our recent research on Langevin dynamics with self-harmonic drift and classical spins has revealed a strong connection to ``Stochastic Localization," a concept attracting attention in the fields of mathematical geometry and data science.
Roughly speaking, the former can be viewed as a version of the latter in which the $n$-dimensional Euclidean space - the domain of the probability distribution function appearing in the latter - is ``compactified'' to the surface of an $n$-dimensional sphere.
By comparing the two frameworks, we argue that this compactification is a consequence of requiring the drift to be autonomous. The relationship between the two approaches may expand the scope of Stochastic Localization.
}

\section{Introduction}
\label{sec:intro}
Since the introduction of the martingale concept into stochastic thermodynamics by Ch\'etrite and Gupta \cite{martingale-Gupta2011} and by Neri, Rold\'an and J\"ulicher\cite{Roldan-prX2017}, the theory has come to be recognized as a powerful tool in non-equilibrium physics. Generic routes for constructing martingales include at the very least \cite{review300AdvInPhys}; (a) path probability ratios, (b) Doob martingales, or, conditional expectation with increasing filtrations, (c) harmonic functions, and (d) Dol\'eans-Dade type functionals. Stochastic thermodynamics employs the type corresponding to (a). Subsequently, type (b) martingales were identified in physical processes such as the sequential fixing of Ising spins on a complete graph \cite{PQ-CM-KS-2020} and the sequential removal of edges in Markov transition networks \cite{PQ-CM-KS-2024}, which we called ensemble ``progressive quenching." In the course of preparing a comprehensive review \cite{review300AdvInPhys} integrating these concepts, it became clear that types (a) and (b) share a common element behind: the ``tower property" of conditional probabilities. Furthermore, we have recently investigated harmonic-function-type martingales (c) as will be described more below \cite{PQL2024,PQL2025}; meanwhile, in the fields of mathematics and data science, ``stochastic localization" \cite{Eldan2013} -- discussed below also --represents a form that extends or ``tilts" the Dol\'eans-Dade type martingale (d).
\\  

In the aforementioned story, {martingale} has primarily been utilized as a mathematical tool for statistical analysis. This is likely due to its inherent affinity with characteristics of physics such as causality and memory. In recent years, however, we have become interested in viewing {martingale} from a slightly different perspective and have investigated cases where {martingale} itself intrinsically embodies physical aspects. Specifically, we examined Langevin equations in which the drift  (the induced stochastic process) is also {martingale} associated with the process generated by the very Langevin equation, that is, ``self-harmonic" \cite{PQL2024}. In this case, the drift  can be expressed as the canonical average of a classical spin under the field that obey the Langevin equation. Furthermore, it has been shown that the canonical probability density of the spin under the aforementioned field also exhibits a {martingale} functional process  \cite{PQL2025}. 
For simplicity, we shall call our previous framework \cite{PQL2024,PQL2025} by Autonomous Self-hamonic Drift.\\

This phenomenon involving classical spins is interesting, but the origin of the spin has remained a mystery. Quite recently, however, we learned that R. Eldan \cite{Eldan2013} had devised a similar phenomenon in the field of mathematics, terming it ``Stochastic Localization" (hereafter SL). 
While this formalism does not contain a spin space but unbounded Euclidean one, our works share both the asymptotic convergence of the stochastic density to a $\delta$-function (``localization'') and the identity in law between the ensemble of the localized (orientational) position with the initial density \cite{PQL2024,PQL2025}. 
Upon further scrutiny (see below) we reached a perspective about the emergence of the spin space as a compactification of $n$-dimensional Euclidean space, $\mathbb{R}^n,$ to the hypersphere, $S^{n-1},$ of co-dimension 1, essentially because of the autonomous requirement on the drift.\\

 \null{Incidentally, during the preparation of the present note, there appeared a work on the arXiv \cite{UCR-arXiv-StLoc} that considers the SL with the stochastic density on hypersphere $S^{n-1}.$ This is done independently from our works \cite{PQL2024,PQL2025}, and is the first approach from SL to the spin space.}\footnote{\null{\protect\cite{UCR-arXiv-StLoc} calls their framework Discrete Stochastic Localization (DSL) aiming at a continuous-state framework for discrete sequence generation.}} 
 
 The organisation of the paper is the following: We summarize Autonomous Self-hamonic Drift in {\S}II and then describe in {\S}III those aspects of SL  \cite{Eldan2013} that we find correspondence in our framework. Subsequently, in {\S}IV, we present a table comparing the previous two sections (Table \ref{tab:compare}). 
 We there discuss, on the one hand, 
 how the imposition of autonomous drift leads to the compactification of the domain of the stochastic density whose barycenter is the drift\footnote{\null{cf. In \protect\cite{UCR-arXiv-StLoc} the autonomous drift is an outcome of the restriction of domain from $\mathbb{R}^n$ to $S^{n-1}.$}}, and,  
 on the other hand, the compactness of the support of the stochastic density allows a type of transferring the initial "tilt", which has no counterpart in the original SL.

\section{R\'esum\'e of {Autonomous self-harmonic drift}}
We summarise below these aspects in \cite{PQL2024,PQL2025} which we can compare with SL.
Throughout the following discussion, we should be careful in distinguishing between two statistical levels: the ``lower level" associated with $x$ and $\hat{x}$, as the variables of stochastic densities, and the ``upper level" associated with ``field" $\theta_t$ and ``densities" $p_t$ and $p_{\theta_t},$ being driven by the Wiener noise, $dW_t.$ (See  \cite{LNP} \S 6.2 for more discussion about the system having different levels of random variables.\footnote{In that literature, $p_{\theta_t}(\hat{x})$ may describe the distribution of polar segments in an ideal chain, $\{\hat{x}\},$ being submitted under the uniform field $\theta_t,$ and $a(\theta_t)$ may represent the scaled mean end-to-end distance of the chain.})
For instance, $a(\theta)$ is the drift in the stochastic evolution of $\theta_t$ but at the same time the barycenter of the density $p_\theta (\hat{x}).$  Even though the initial density is set to be symmetric ($p_0(-\hat{x}) = p_0(\hat{x})$), this is not the case for $p_{\theta_t} (\hat{x})$ because the evolution of $\theta_t$ introduces a "tilt" in the density $p_{\theta_t}$ as function of $\hat{x}.$

The main equations are copied to the left column of TABLE.\ref{tab:compare}. The numbering (i)-(vi) are in parallel with those in {\S}.III. See Appendix \ref{app:B} for the symbols. 

{\null}{(i)} In $\mathbb{R}^n$ space $(W_t)_{t\ge 0}$ is a Wiener process and $(\theta_t)_{t\ge 0}$ is a stochastic process generated by the SDE: 
\beq  \label{eq:dtheta}
d\theta_t=a(\theta_t) dt + dW_t.
\eeq
The initial value $\theta_0$ is specified.

{\null}{(ii)}
The drift $a(\theta),$ is imposed to be self-harmonic, that is, the 
process $(a(\theta_t))_{t\ge 0}$ is {martingale} with respect to the process $(\theta_t)_{t\ge t}$ whose drift is $a(\theta)$ itself. 
\null{Furthermore, $a(\theta)$ should be autonomous, i.e. $a(\theta)$ does not explicitly depend on time.}

As an outcome, the drift $a(\theta)$ is the canonical average of a classical spin on the hypersphere, $ S^{n-1}:=\{\hat{x}\in \mathbb{R}^n |\, \|\hat{x}\|=1\},$ which we denote by 
$a(\theta)=\langle \hat{x} \rangle^{{\null}}_{\theta}.$\footnote{For $n=3$ and constant $U(\hat{x})$ the module of $a(\theta)$ is called a Langevin function \protect\cite{langevinfunction}, $\|a(\theta)\|=\mathcal{L}(\theta):= \coth \theta-{1/\theta}.$}
(It means that $a(\theta) \not\in S^{n-1}$ for $t<\infty.$)
\null{cf. The authors of \cite{UCR-arXiv-StLoc} reached this form by limiting the $\mathbb{R}^n$ space of $x$ to its co-dimension 1 subset, $S^{n-1}.$}

{\null}{(iii)}
The canonical density that gives $a(\theta)$ should have the support $S^{n-1}(\subset \mathbb{R}^n)$ and reads:
\beq   \label{eq:px}
p^{{\rm \null}}_\theta(\hat{x}):= e^{\langle \theta,\hat{x}\rangle\null{-U(\hat{x})}
}/Z(\theta)
\eeq
with $Z(\theta)=\oint e^{\langle \theta,\hat{x}\rangle\null{-U(\hat{x})}} d\Omega_{\hat{x}}.$ 
Here $\langle \alpha,\beta\rangle$ with $\alpha,\beta\in\mathbb{R}^n$ 
means to take a scalar product, and
$\oint d\Omega_{\hat{x}}$ is the surface integral over $S^{n-1}.$ 
The classical mean spin $a(\theta)$ then reads
$a(\theta)=\langle \hat{x}\rangle_\theta= \oint \hat{x}p_\theta(\hat{x})d\Omega_{\hat{x}}.$

\null{The possibility of inhomogeneous background measure, $p_0(\hat{x}):=e^{-U(\hat{x})},$ has been introduced by \cite{UCR-arXiv-StLoc} as a natural extension of Eldan's original framework \cite{Eldan2013}.} The precedent works \cite{PQL2024,PQL2025} have not allowed for this while the calculation of self-harmonic condition remains basically the same.

It turns out that $p^{{\rm \null}}_{\theta_t}(\hat{x})$ satisfies the functional SDE;
\beq  \label{eq:dpx}
dp^{{\rm \null}}_{\theta_t}(\hat{x})= p^{{\rm \null}}_{\theta_t}(\hat{x}) \langle \hat{x}-a(\theta_t), dW_t\rangle ,
\eeq
where $(W_t)_{t\ge 0}$ is what generated $(\theta_t)_{t\ge 0}$ in Eq.(\ref{eq:dtheta}) above.
In Eq.(\ref{eq:dpx}) the factor $dW_t$ means that for each $\hat{x}$ the process $(p_{\theta_t}(\hat{x}))_{t\ge 0}$  is {martingale}, 
\beq \label{eq:martingale}
\mathbb{E}[p^{{\rm \null}}_{\theta_t}(\hat{x}) | (\theta_s)_{0\le s\le u}
]=p^{{\rm \null}}_{\theta_u}(\hat{x}),
\quad t\ge u\ge 0,
\eeq
while the factor $p^{{\rm \null}}_{\theta_t}(\hat{x}) (\hat{x}-a(\theta_t))$ in Eq.(\ref{eq:dpx}) guarantees the normalization of $p^{{\null}}_{\theta_t}(.)$. 
Here $\mathbb{E}$ is a (conditional) expectation over path ensemble.

{\null}{(iv)} 
For $t\to\infty$ the field $\theta_t$ obeying (\ref{eq:dtheta}) diverges as
$\theta_t\simeq t\, {{\hat{x}^*}}+W_t \quad (t {\to\infty}).$\footnote{We do not exclude the analogous relationship to \S III (iv) \protect\cite{Klartag-Putterman2021} in SL for the moment. Neither Bayesian analyses nor numerical tests are easy problems.}
It is the growing field $\theta_t$ that eventually localizes {\it and} orients the spin at some $\hat{x}^* (=a(\theta_{\infty}))\in S^{n-1}$:
\beq    \label{eq:p-localized} 
p^{{\rm \null}}_{\theta_t}(\hat{x})\simeq \delta(\hat{x}-{\hat{x}^*})\quad (t\to\infty),
\eeq
where $\delta(\hat{x}-{\hat{x}^*})$ should be normalized on $S^{n-1}.$

{\null}{(v)}
By the martingality and localization we have $\mathbb{E}[\delta(\hat{x}-{{\hat{x}^*}})]=p^{{\rm \null}}_{\theta_0}(\hat{x}),$ i.e., the law of stochastic localization:
\beq \label{eq:StLog}
\{\hat{x}^*\}
 {\sim}  p^{{\rm \null}}_{\theta_0}(\hat{x}^*)\quad \mbox{(in law)}
\eeq

{\null}{(vi)} From data transfer point of view, we can have two approaches and their mixture in principle:
1) Starting with $\theta_0=0$ the density of $\{x^*\}$ gives the initial background density, $e^{-U(x^*)},$ in the spirit of the sampling through SL \cite{Eldan2013}.
2) 
The initial field $\theta_0$ is recovered by fitting the density of $\{x^*\}$ with $p_{\theta_0}(x^*)$ \cite{PQL2024}.
\\

Eq.(\ref{eq:dtheta}) with the result $a(\theta)=\langle\hat{x}\rangle_\theta$ 
 allows us to interpret the Langevin process with autonomous martingale drift in the following manner: 
(a) In the presence of already frozen magnetization $\theta_t,$
(b) a new fragment of (magnetic) spin $d\theta_t$ is created and fixed
during the time $dt,$ which is stochastically given as the sum of mean 
increment proportional to the equilibrium polarisation, $a(\theta_t),$ and the Wiener noise $dW_t,$ then  
(c) the newly frozen spin $d\theta_t$ is added to the existing part $\theta_t,$ and so on.
This is similar to the ``progressive quenching'' in \S I \cite{PQ-CM-KS-2020}.

\section{R\'esum\'e of original Stochastic Localization }
We follow the original description {Stochastic Localization} of  Eldan \cite{Eldan2013} but also refer a simplified version by Lee and Vempala \cite{Lee-Vempala2017,Lee-Vempala2024}. 
{Reciprocally to \S II, we summarise below those aspects in the stochastic localisation that we can compare with Autonomous Self-harmonic Drift. We, therefore, will not address the high-dimensional geometry or the sampling in a very high dimensional space.}

The main equations below are copied to the right column of TABLE.\ref{tab:compare}. The enumeration (i)-(iv) below correspond to those in \S.II.

{\null}{(i)} In $\mathbb{R}^n$ space $(W_t)_{t\ge 0}$ is a Wiener process and 
   $(\theta_t)_{t\ge 0}$ is a stochastic process generated by the SDE
\beq    \label{eq:dtheta1}
d\theta_t=a(t,\theta_t) dt+dW_t 
\eeq
with the initial condition,  $\theta_0=0.$

{\null}{(ii)} 
The drift $a(t,\theta)$ is imposed to be the barycenter of a distribution $p_{t,\theta}(x)$ defined on the space $\{x\}=\mathbb{R}^n.$ It reads $a(t,\theta)=\langle x\rangle_{t,\theta}\equiv \int_{\mathbb{R}^n}x \,p_{t,\theta}(x) dx.$   Evidently $a(t,\theta)\in \mathbb{R}^n.$ {Here $a(t,\theta)$ depends explicitly on time, $t.$}
That $a(t,\theta_t)$ is {martingale}, or self-harmonic, is an outcome of the setting of $p_{t,\theta_t}(x)$ (see below). 

{\null}{(iii)}
The form of distribution $p_{t,\theta}(x)$ whose barycenter is $a(t,\theta),$ is imposed:
\beq      \label{eq:px1}
p_{t,\theta}(x)\equiv e^{\langle \theta,x\rangle-({t}/{2})\|x\|^2} p_0(x)/{Z(t,\theta)}, 
\eeq
with $Z(t,\theta)=\int_{\mathbb{R}^n} e^{\langle \theta,x\rangle-({t}/{2})\|x\|^2} p_0(x) dx.$\\
The stochastic measure $p_t(x)$ is defined as $p_t(x):=p_{t,\theta_t}(x)$ for $t\ge 0.$

It turns out that $p_t(x)$ satisfies the functional SDE;
\beq  \label{eq:dpx1}
 dp_t(x)= p_t(x) \langle x-a(t,\theta_t), dW_t\rangle, 
\eeq
where $(W_t)_{t\ge 0}$ is what generated $(\theta_t)_{t\ge 0}$ in the above.
The process  $(p_{t}({x}))_{t\ge 0},$ with ${x}$ being fixed, is, therefore, {martingale}: 
\beq   \label{eq:martingale1}
\mathbb{E}[p_{t}({x})| (\theta_s)_{0\le s\le u}]=p_{u}({x}), \quad t\ge u\ge 0.
\eeq

{\null}{(iv)}
It has been shown \cite{Klartag-Putterman2021} that the process $(\theta_t)_{t\ge 0}$ obeying the above SDE coincides {\it in law} with $(\tilde{\theta}_t)_{t\ge 0}$ defined by
$ \tilde{\theta}_t= t\, {x^*}+\tilde{W}_t, $
 where $x^* $ obeys ${p_0(x^*)}$ independently of  another Wiener process, $(\tilde{W}_t)_{t\ge 0}.$

For $t\to\infty$ it is the sharpening Gaussian exponent, $-(t/2)\|x\|^2,$ that makes $p_t(x)$ localise around $x^*\simeq \theta_t/t $ : 
\beq    \label{eq:p-localized1} 
p_{t}({x})\simeq \delta({x}-{x^*}) \quad (t\to\infty).
\eeq

{\null}{(v)}
By the martingality and localization we have $\mathbb{E}[\delta({x}-{x^*})]=p_{0}({x}),$ i.e., the law of stochastic localization:
\beq   \label{eq:StLoc1}
\{{x}^*\}  {\sim}  p_{0}(x^*)\quad \mbox{(in law)}
\eeq
\mbox{ }

{\null}{(vi)}   From data transfer point of view, Eq.(\ref{eq:StLoc1}) above tells how the initial data $p_0$ is statistically conveyed to the ensemble of asymptotically localized densities, $\{x^*\}.$
 Unlike with the hypersphere $S^{n-1},$ which is bounded, we cannot start by a flat density, $p_0=\mbox{const.},$ on the unbounded support, $\mathbb{R}^n.$
(By contrast, if the initial density $p_0(x)$ is localized, say $\delta(x-x_i),$
then it evolves unchanged as a stable fixed point function of (\ref{eq:dpx1}).)

\section{\protect{Comparison of the two frameworks}}

To relook at Autonomous Self-harmonic Drift in the light of SL, we compared the summaries in \S II and \S III in TABLE \ref{tab:compare}. In so doing we put priority to keep the correspondence of the equations, rather than the logical order of derivation. To compensate this ignorance, 
 we indicated by  $\bullet$ and $\circ$ to mean, respectively, the imposed/defined properties and the obtained/derived properties.\footnote{\null{ cf. In Table 5 of \protect\cite{UCR-arXiv-StLoc} the authors show a pedagogical comparison between their model (DSL) and prior Continuous Diffusion Language Models.}}

In TABLE \ref{tab:compare} we first notice that, although some properties are imposed ($\bullet$) on one side while they are derived ($\circ$) on the other side, the principal order of logic from the SDE for the field, $\theta_t,$ to the SDE for the density, $p_t,$ is common.
\null 
One could start with the SDE for the density $p_t(x)$ (eqs.(\ref{eq:px}) and (\ref{eq:px1})), but then all the components $x$ or $\hat{x}$ should couple with each other under a single noise $dW_t,$ which would seem to be physically unconceivable, as noticed in Chap.6 of \cite{Lehec2024}.

\null{While the paper \cite{UCR-arXiv-StLoc} adopted as the domain of stochastic densities a ball $S^{n-1} (\subset \mathbb{R}^n),$} we have observed that a spin space can emerge by simply imposing that the drift be autonomous - in addition to the martingale property. Let us consider the implications of this. The martingale property of the drift, $a,$ implies that its infinitesimal evolution $da$ from $t$ to $t+dt$ is neutral (zero) on average given the history up to time $t$; 
Applying It\^{o}'s formula to this condition requires:
\beq \label{eq:a-mtgl}
\partial_t a+ \langle a, \nabla_\theta\rangle a+\frac{1}{2}\Delta_\theta \,a=0.
\eeq 
By an analogy to statistical thermodynamics, we assume the gradient-type drift and use the Riccati/Cole-Hopf transformation, $a=\nabla_\theta \log Z,$ to rewrite the last two terms on the l.h.s. of Eq.(\ref{eq:a-mtgl}). This operation yields $(1/2)\nabla_\theta (\Delta_\theta Z/Z).$
 If the drift $a$ is allowed to depend explicitly on time - as in the SL case - the first term $\partial_t a(t,\theta)$ cancels this out. This relates to the sharpening Gaussian factor, $e^{-(t/2)\|x\|^2},$ in the probability density, $p_{t,\theta},$ defined in Eq.(\ref{eq:px1}).\footnote{The factor $e^{-(t/2)\|x\|^2}$ is inherited from Dol\'eans-Dade exponential 
 (\protect\cite{Lehec-HDR},\S 4.1): When $a(t,\theta_t)=0,$  Eq.(\ref{eq:dpx1}) becomes the SDE; $dG_t(x)=G_t(x)$ $ \langle x,dW_t\rangle,$ being solved as $G_t(x)=G_0(x)e^{\langle x,W_t\rangle-(t/2)\|x\|^2 }.$} However, when $a$ is autonomous, the term $\partial_t a(\theta)$ vanishes and, therefore, the remaining terms must cancel on their own. The form, $Z(\theta)=\oint_{\|\hat{x}\|=1} e^{\langle \hat{x},\theta\rangle-U(\hat{x})} d\Omega_{\hat{x}},$  achieves this by the relation, $\Delta_\theta Z(\theta)=Z(\theta)$ because of the normalization of $\hat{x}.$
 This is our conceptual picture of how spin space emerges from autonomous drift. The localization of the probability density $p_{\theta_t}(\hat{x})$ is realized without the synchronized sharpening Gaussian factor since the martingale property of $a$ lets the field $\theta_t$ to strengthen ballistically ($\sim t$).  This field then plays a double-role; orienting the spin along the field axis and   suppressing its off-axis fluctuations. (If we need standardized convergences of $a(\theta_t)$ or $p_{\theta_t},$ we might resort to
 the Optional Stopping Theorem (OST) of martingale processes, see for example  Sec.4.1.5 of \cite{review300AdvInPhys}.
\hspace{-10mm} 
{\small
\begin{table}[h] 
\caption{{Comparison between Autonomous Self-harmonic Drift (left) and original Stochastic Localization (right).}}
\begin{center}
\begin{tabular}{|cl | l |}
\hline
 &
Autonomous self-harmonic drift \protect\cite{PQL2024,PQL2025}
& 
Original stochastic localization \protect\cite{Eldan2013,Lee-Vempala2017,Lee-Vempala2024}
\cr \hline  \hline
(i)& 
$(W_t)_{t\ge 0}$ :  Wiener process in $\mathbb{R}^n$ 
&
$(W_t)_{t\ge 0}$ :  Wiener process in $\mathbb{R}^n$ 
\cr \null &
$(\theta_t)_{t\ge 0}$: ``field'' process in $\mathbb{R}^n$ 
&
$(\theta_t)_{t\ge 0}$: ``field'' process in $\mathbb{R}^n$ 
\cr \null &
$\bullet$
$d\theta_t=a(\theta_t) dt + dW_t$ (\ref{eq:dtheta}). \quad $\theta_0$: specified.
&
$\bullet$
$d\theta_t=a(t,\theta_t) dt+dW_t$ (\ref{eq:dtheta1}). \quad $\theta_0:=0.$

\cr \hline
(ii)& 
Drift $a(\theta)$: 
& 
Drift $a(t,\theta)$: 
\null \cr &
$\bullet$ 1) $a(\theta_t)$ : {martingale} 
&
$\circ$ 1) $a(t,\theta_t)$ : {martingale} 
\null \cr &
$\bullet$ 2) autonomous
& 
$\bullet$ 2) {\it non}-autonomous
\null \cr  &
$\circ$  3)  $a(\theta)=\langle \hat{x} \rangle^{{\null}}_{\theta}$ ($a(\theta)\!\not\in \! S^{n-1} \!\mbox{ for }\! \|\theta\| \! <\! \infty$)
&
$\bullet$ 3)  $a(t,\theta)=\langle x\rangle_{t,\theta}$ 
\cr \hline
(iii)& 
$\circ$ Density: $p^{{\rm \null}}_\theta(\hat{x}):= \frac{e^{\langle \theta,\hat{x}\rangle-U(\hat{x})} }{Z(\theta)}$ \hfill  (\ref{eq:px})
&
$\bullet$ Density: $p_{t,\theta}(x)\equiv \frac{e^{\langle \theta,x\rangle-({t}/{2})\|x\|^2} p_0(x)}{{Z(t,\theta)}}$ \hfill (\ref{eq:px1})
\null \cr &
$\circ$ Support: $\hat{x}\in S^{n-1}$  ($\subset \mathbb{R}^n$: Compactify)
& 
$\bullet$ Support: $x\in \mathbb{R}^n$
\null \cr &
$\circ$ SDE: 
$dp^{{\rm \null}}_{\theta_t}(\hat{x})= p^{{\rm \null}}_{\theta_t}(\hat{x}) \langle \hat{x}-a(\theta_t), dW_t\rangle $  \hfill (\ref{eq:dpx})
&
$\circ$ SDE: 
$ dp_t(x)= p_t(x) \langle x-a(t,\theta_t), dW_t\rangle$  \hfill (\ref{eq:dpx1})
\null \cr &
$\circ$ Martingale: $\mathbb{E}[p^{{\rm \null}}_{\theta_t}(\hat{x}) | (\theta_s)_{0\le s\le u}]=p^{{\rm \null}}_{\theta_u}(\hat{x}) $  \hfill (\ref{eq:martingale})
&
$\circ$ Martingale: $\mathbb{E}[p_{t}({x}) | (\theta_s)_{0\le s\le u}]=p_{u}({x})$  \hfill (\ref{eq:martingale1})
 \cr \hline
 (iv)& 
 $\circ$ Asymptot: $\theta_t\simeq t\, {{\hat{x}^*}}+W_t \, (t {\to\infty}).$
&
$\circ$ Asymptot: $\theta_t\sim t x^* +\tilde{W}_t$ (in law) 
\null \cr 
&
$\circ$ Localization: 
$p^{{\rm \null}}_{\theta_t}(\hat{x})\simeq \delta(\hat{x}-{\hat{x}^*})\, (t\to\infty).$  \hfill (\ref{eq:p-localized})
&
$\bullet$ Localization: 
$p_{t}({x})\simeq \delta({x}-{x^*}) \, (t\to\infty).$ \hfill (\ref{eq:p-localized1})
\null \cr &
$\circ$ Localizing factor: Growing, $\theta_t,$ upon spin
&
$\bullet$ Localizing factor: Sharpening  $e^{-(t/2)\|x\|^2}$
\cr \hline
(v)& 
$\circ$ Stochastic localization as (iii)$\wedge$(iv) 
&
$\circ$ Stochastic localization as (iii)$\wedge$(iv)  
\cr
\null &
\quad $\to$  $\mathbb{E}[\delta(\hat{x}-{{\hat{x}^*}})]=p^{{\rm \null}}_{\theta_0}(\hat{x}),$ or $\hat{x}^* \sim p^{{\rm \null}}_{\theta_0}(\hat{x}).$  \hfill (\ref{eq:StLog})
&
\quad $\to$  $\mathbb{E}[\delta(x-x^*)]=p_0(x),$ or $ x^*\sim p_0(x^*).$  \hfill (\ref{eq:StLoc1})
  \cr \hline
(vi)& Information in $\{\delta(\hat{x}-\hat{x}^*)\}$:
& Information in $\{\delta({x}- {x}^*)\}$:
\cr &
$\circ$ $\hat{x}^* \sim e^{-U(\hat{x}^*)}$ \& $\theta_0=0$ to sample $e^{-U(\hat{x})}$
& $\circ$  $x^* \sim p_0(x^*)$ \& $\theta_0=0$ to sample $p_0(x)$
\cr & 
$\circ$ $\hat{x}^*\sim p_{\theta_0}(\hat{x}^*)$ to find $\theta_0$
& $\times$  \mbox{No counterpart}

\cr \hline
\end{tabular}
\end{center}
\label{tab:compare}
\end{table}%
}

\section*{Concluding discussion}
  Stochastic Localization is an important advancement in high-dimensional geometry approached by inequalities, but also as complementary approach to the diffusion models (see, for example,  the Lecture Note \cite{Lehec2024}).

 The relationship between the SL and Autonomous Self-harmonic Drift may expand the scope of the former. Spin emerged as a result of compactification of $\mathbb{R}^n$ space for $x.$ It is possible to develop the SL theory having the stochastic density on the cylindrical space like $\mathbb{R}^{n_1}\otimes S^{n_2-1}
( \subset \mathbb{R}^{n_1+n_2}).$

The martingale relation (\ref{eq:martingale}) for the case $u=0,$ i.e.,  $\mathbb{E}[p^{{\rm \null}}_{\theta_t}(\hat{x}) ]=p^{{\rm \null}}_{\theta_0}(\hat{x})$ holds strictly even for finite $t$. Therefore, if sufficient samples of $\{\theta_t\}$  are obtained starting from a specific value of $\theta_0,$ it is possible in principle to recover $\theta_0$ using this relation, even when direct observation of $\theta_0$ is difficult.

 The Autonomous Self-harmonic Drift brings also
the physical picture of ``progressive quenching," where $\theta$ and $\hat{x}$ can be regarded not merely as variables corresponding to the upper and lower levels of statistics, respectively, but also as a pair of thermodynamically conjugate variables. The martingale time evolution of $a = \langle \hat{x} \rangle_\theta$ can be written as $da(\theta_t)=(\langle \hat{x}\hat{x}\rangle_\theta- \langle \hat{x}\rangle_\theta \langle \hat{x}\rangle_\theta)\cdot dW_t,$ with coefficients 
similar to the fluctuation-response relation, see for example \cite{Oono-book} \S 24.4-6. 
 \\


\section*{Acknowledgements}
We acknowledge \null{\'{E}dgar Rold\'{a}n for his early pointing out of the relevance of our previous works to the Boltzmann machine, } and Pierfrancesco Urbani for having brought us to the field of Stochastic Localization. \null{We also thank the authors of \cite{UCR-arXiv-StLoc} for the constructive communications about their preprint.}

\begin{appendix}
\section{Symbols in different articles}\label{app:B}
For the purpose of later comparison with Stochastic Localisation (SL) \cite{Eldan2013}, the symbols used in our papers \cite{PQL2024,PQL2025} are readapted to those more common in SL, for example, in the lecture note, \cite{Lehec2024},  Chap.6 :\\
\indent 
$\{X_t, \hat{S}, a\}\mapsto \{\theta_t,\hat{x}, a_\theta \},$ from \cite{PQL2024,PQL2025}, 
\\
 while $W_t$ for the Wiener process remains the same. Also the main mapping from the notations in \cite{UCR-arXiv-StLoc} to those of the present paper is: \\
\indent 
$\{z,\hat{x},x, P\}\mapsto \{\theta, a_\theta, \hat{x}, e^{-U}\},$ from \cite{UCR-arXiv-StLoc}.

\end{appendix}

\paragraph{Funding information}

\bibliographystyle{SciPost_bibstyle}
\bibliography{ken_LNP_sar}

\end{document}